\documentclass[aps,prc,twocolumn,floatfix]{revtex4}

\usepackage{graphics}
\usepackage{epsfig}

\begin{document}
\bibliographystyle{/home/anogga/.TeX/apsrev}

\title{The three-nucleon bound state using realistic potential models}

\author{{A.~Nogga$^1$ \email{anogga@physics.arizona.edu}, 
         A.~Kievsky$^{2,3}$, 
         H.~Kamada$^4$, 
         W.~Gl\"ockle$^5$,
         L.E.~Marcucci$^{2,3}$,
         S.~Rosati$^{2,3}$,
         M.~Viviani$^{2,3}$}}

\affiliation{
$^1$Department of Physics, University of Arizona, Tucson,
    Arizona 85721, USA \\
$^2$Istituto Nazionale di Fisica Nucleare, Via Buonarroti 2, 
    56100 Pisa, Italy \\
$^3$Dipartimento di Fisica, Universita' di Pisa, Via Buonarroti 2,
    56100 Pisa, Italy \\
$^4$Department of Physics, Faculty of Engineering, 
    Kyushu Institute of Technology, Kitakyushu 804-8550, Japan \\
$^5$Institut f\"ur theoretische Physik II, Ruhr-Universit\"at Bochum,
    D-44780 Bochum, Germany }

\begin{abstract}
The bound states of $^3$H and $^3$He have been calculated using the
Argonne $v_{18}$ plus the Urbana three-nucleon potential. The 
isospin $T=3/2$ state have been included in the calculations as well
as the $n$-$p$ mass difference. The  $^3$H-$^3$He mass difference has 
been evaluated through the charge dependent terms explicitly included 
in the two-body potential. The calculations have been performed
using two different methods: the solution of the Faddeev equations
in momentum space and the expansion on the correlated
hyperspherical harmonic basis. The results are in agreement within $0.1$\%
and can be used as benchmark tests.
Results for the CD-Bonn interaction are also presented. It is shown
that the $^3$H and $^3$He binding energy difference can be predicted 
model independently. 
\end{abstract}

\maketitle

In the last years great efforts have been made to improve the
description of the nucleon-nucleon ($NN$) interaction. A new
generation of potentials including explicitly 
charge independence and charge symmetry
breaking (CIB,CSB) terms appeared.
These interactions describe the $NN$ scattering 
data below $T_{lab}=300$~MeV with a nearly perfect 
$\chi^2$/datum$\approx 1$.
The CD-Bonn \cite{machleidt96} and Argonne~$v_{18}$ (AV18) \cite{wiringa95} 
interactions also provide a neutron-neutron ($nn$) force, 
which has been adjusted to the experimental 
$nn$ scattering length, whereas the 
Nijmegen interactions \cite{stoks94} are fitted only to proton-proton
and proton-neutron data. 
Recently, the CD-Bonn potential has been updated to CD-Bonn~2000 
\cite{machleidt01a}. In this paper we only present results for the 
AV18 and CD-Bonn~2000 interactions. Both are quite different from each 
other in their functional form, but their description of the $NN$ data is 
almost equally accurate.  Therefore a comparison of the results 
will give insights into the model dependence of our understanding of the 
three-nucleon ($3N$) bound states.

Following for example the notation of Ref.~\cite{wiringa95}, 
all these $NN$ potentials can be put in the general form
\begin{equation}
   v(NN)=v^{EM}(NN)+v^\pi(NN)+v^R(NN)\ .
\end{equation}

The short range part $v^R(NN)$ of all of these interactions
includes a certain number of parameters (around 40), 
which are determined by a fitting procedure to the $NN$ scattering data and
the deuteron binding energy (BE), whereas the long range part 
is represented by the one-pion-exchange potential $v^\pi(NN)$
and an electromagnetic part $v^{EM}(NN)$. 

For AV18, $v^{EM}(pp)$ 
consists of the one- and two-photon Coulomb terms plus the
Darwin-Foldy term, vacuum polarization and magnetic moment interactions.
The $v^{EM}(np)$ interaction includes a Coulomb term due to the neutron charge
distribution in addition to the magnetic moment interaction. Finally,
$v^{EM}(nn)$ is given by the magnetic moment interaction only. All
these terms take into account the finite size of the nucleon charge
distributions. The $v^{EM}(NN)$ for CD-Bonn is much simpler: 
$v^{EM}(pp)$ is given by the Coulomb force of point protons, whereas 
$v^{EM}(np)=v^{EM}(nn)=0$.

As it is well known, when these interactions are used to describe
the $3N$ bound state, an underbinding  
of about $0.5$ MeV to $0.9$ MeV depending on the model is obtained
(see for example Ref.\cite{nogga00}). 
The local potentials lead to less binding than 
the non-local ones, a characteristic related to the bigger $D$-state
probability predicted  for the deuteron. 
Hence, it seems to be not
possible to describe the $A>2$ systems without the inclusion
of three-nucleon interaction (TNI) terms in the nuclear Hamiltonian. 
Several TNI models have been studied in the literature mostly based
on the exchange of two pions with an intermediate $\Delta$ 
excitation (for a recent review see Ref.~\cite{carlson98}). 
These interactions include a certain
number of parameters not completely determined by theory, therefore
some of them can be  used to reproduce, for example, the triton BE.

In the following we show BE  results 
for $^3$H and $^3$He.
The AV18 interaction in conjunction with the Urbana IX
(UIX) TNI~\cite{pudliner95} has recently been discussed by the Argonne 
group in a  study of several bound states
with  $3\le A \le 8$ \cite{pieper01}. 
Here we perform a detailed calculation of the
$A=3$ system including total isospin states $T=1/2$ and $3/2$. In addition,
particular attention will be given to the BE difference 
$D=B(^3{\rm H})-B(^3{\rm He})$ as a test of the CSB terms present
in the interaction. The experimental value of this
quantity is $764$ keV, from which only $85$\% correspond to
the standard Coulomb potential \cite{friar87,wu90,wu93}. 
The remaining $15$\% should come from other CSB terms.
A previous analysis of the contributions to $D$ has been performed
a decade ago~\cite{wu90,wu93} before the construction of the new 
interactions, which include for the first time the 
CSB terms in the fit to the $NN$ data. Therefore a reanalysis 
is in order and might remove uncertainties due to an 
inaccurate description of the $NN$ data.

\begin{table*}[tbh]

\begin{tabular}{l|ccc|cccc}
Hamiltonian         & $|E|$       &   $B$    &   $T$   &  $P_{S'}$ &  $P_P$ &  
$P_D$ &  $P_{T=3/2}$ \cr
\hline
AV18 ($T=1/2$)       &  ---      &  7.618   & 46.714  & 1.295  & 0.066  & 8.510  &   ---        \cr
AV18 ($T=1/2,3/2$)   &  ---      &  7.624   & 46.727  & 1.293  & 0.066  & 8.510  &  0.0025      \cr
AV18+UIX($T=1/2$)    &  ---      &  8.474   & 51.262  & 1.055  & 0.135  & 9.301  &   ---        \cr
AV18+UIX($T=1/2,3/2$)&  ---      &  8.479   & 51.275  & 1.054  & 0.135  & 9.301  &  0.0025      \cr
\hline
AV18 ($T=1/2$)       &  7.622    &  7.616   & 46.73   & 1.290  & 0.066  & 8.510  &   ---        \cr
AV18 ($T=1/2,3/2$)   &  7.621    &  7.621   & 46.73   & 1.291  & 0.066  & 8.510  &  0.0025      \cr
AV18+UIX($T=1/2$)    &  8.477    &  8.470   & 51.28   & 1.051  & 0.135  & 9.302  &   ---        \cr
AV18+UIX($T=1/2,3/2$)&  8.476    &  8.476   & 51.28   & 1.052  & 0.135  & 9.302  &  0.0025      \cr
\end{tabular}

\caption{$^3$H BE $B$, mean value of the kinetic energy $T$,
         $S'$-, $P$- and $D$-probabilities and 
         the probability of the $T=3/2$ state. 
The Pisa results are displayed in the first four rows. The last four rows
         show the Bochum results, in this case the modulus of the
eigenvalue $E$ is also given. All energies are given in MeV. The 
         probabilities are given in \%. }

\label{tab:hres}

\end{table*}

\begin{table*}[tbh]

\begin{tabular}{l|ccc|cccc}
Hamiltonian         & $|E|$       &   $B$    &   $T$   &  $P_{S'}$ &  $P_P$ 
&  $P_D$ &  $P_{T=3/2}$ \cr
\hline
AV18 ($T=1/2$)       &  ---      &  6.917   & 45.669  & 1.531  & 0.064  & 8.468  &   ---        \cr
AV18 ($T=1/2,3/2$)   &  ---      &  6.925   & 45.685  & 1.530  & 0.065  & 8.467  & 0.0080       \cr
AV18+UIX($T=1/2$)    &  ---      &  7.742   & 50.194  & 1.242  & 0.131  & 9.249  &   ---        \cr
AV18+UIX($T=1/2,3/2$)&  ---      &  7.750   & 50.211  & 1.242  & 0.132  & 9.248  & 0.0075       \cr
\hline
AV18 ($T=1/2$)       &  6.936    &  6.915   & 45.70   & 1.515  & 0.065  & 8.465  &   ---        \cr
AV18 ($T=1/2,3/2$)   &  6.923    &  6.923   & 45.68   & 1.524  & 0.065  & 8.466  & 0.0081       \cr
AV18+UIX($T=1/2$)    &  7.759    &  7.738   & 50.23   & 1.229  & 0.132  & 9.248  &   ---        \cr
AV18+UIX($T=1/2,3/2$)&  7.746    &  7.746   & 50.21   & 1.235  & 0.132  & 9.248  & 0.0075       \cr
\end{tabular}

\caption{Same as Table~\ref{tab:hres} for $^3$He.}

\label{tab:heres}
\end{table*}

Because the Coulomb energy scales with the BE of $^3$H \cite{friar87},
we need $3N$ Hamilonians, which predict this observable 
accurately. This can be achieved with properly adjusted TNI's.
Then the  calculation of $D$  
requires reliable solutions of the $3N$ Schr\"odinger 
equation including these TNI's. 
For this reason we performed, as a by-product of our analysis,
a benchmark calculation for the bound states of
$^3$H and $^3$He using the AV18+UIX potential model. 
The Bochum group solves the Faddeev equation in momentum
space~\cite{nogga00}, whereas the Pisa group uses a decomposition of the wave
function in pair-correlated hyperspherical basis functions
\cite{kievsky93,kievsky97}. Both methods
were used to find BE's to an accuracy of 2 keV, 
which means an accuracy better than $0.1$\%. 
Such a level of accuracy is nowadays routinely achieved  
for the $3N$ system by several methods using only $NN$ interactions 
\cite{pieper01,suzukilec,kameyama89,navratil98}. 
Here we show that the same
level of accuracy is obtained, when TNI terms are taken into account.

\begin{table}
\begin{tabular}{l|ccc}
 Hamiltonian  & $^3$H &  $^3$He \cr
\hline
AV18          &   6 keV &  -6 keV \cr
AV18+UIX      &   7 keV &  -7 keV \cr
\hline
AV18          &   6 keV &  -6 keV \cr
AV18+UIX      &   7 keV &  -7 keV \cr
\end{tabular}

\caption{Contribution of the proton and neutron mass difference 
 to the $^3$H and $^3$He BE. The Pisa results are 
 displayed in the first two rows. The last two rows show the Bochum results.}
\label{tab:pertmass}
\end{table}

\begin{table}

\begin{tabular}{l|ccc}
Interaction term           &  D       \cr
\hline
nuclear CSB                &  65  keV \cr
point Coulomb              &  677 keV \cr
full Coulomb               &  648 keV \cr
magnetic moment            &  17  keV \cr
$n$-$p$ mass difference    &  14  keV \cr
\hline
total (theory)             &  744 keV \cr
experiment                 &  764 keV \cr
\end{tabular}

\caption{Contributions of the various terms of the interaction to the 
         $^3$H--$^3$He mass difference D. The AV18+UIX potential has been
         used}
\label{tab:csbcontr}
\end{table}

\begin{table}[tp]
  \begin{center}
    \begin{tabular}[t]{l|rr|rr|r}
  & \multicolumn{2}{c|}{$^3$H} & \multicolumn{2}{c|}{$^3$He} &    \cr
                  &  $|E|$  &    $T$   &  $|E|$  &    $T$  & $D$ \cr
\hline
CD-Bonn 2000      &  8.005  &  37.64   &  7.274  & 36.81   &  0.731 \cr
CD-Bonn 2000+TM   &  8.482  &  39.39   &  7.732  & 38.54   &  0.750 \cr
\hline
Exp.              &  8.482  &  ---     &  7.718  & ---     &  0.764   \cr
    \end{tabular}
    \caption{$3N$ BE's $|E|$ for CD-Bonn~2000
             with and without TM-TNI
             compared to the experimental values. 
             Results are shown for $^3$H,$^3$He and their BE 
             difference $D$. 
             Additionally, we show the kinetic energies $T$. 
             All results are given in MeV}
    \label{tab:3nbound}
  \end{center}
\end{table}

\begin{table}[tp]
  \begin{center}
    \begin{tabular}{llll}
   $\lambda$       &    $|E|$  & $D$  &$a_{nn}$   \cr
\hline				       	  
     0.9990        &   8.474          & 742  &-18.75         \cr
     0.9995        &   8.478          & 746  &-18.86         \cr
     1.0000        &   8.482          & 750  &-18.97         \cr
     1.0005        &   8.486          & 754  &-19.08         \cr
     1.0010        &   8.491          & 759  &-19.19         \cr
     1.0020        &   8.499          & 767  &-19.42         \cr
     \end{tabular}
    \caption{Strength factor $\lambda$ for the $^1$S$_0$ $nn$ force, 
             resulting $^3$H BE  $|E|$ in MeV, the BE difference 
             of $^3$He and $^3$H $D$ in keV and  
             $nn$ scattering length $a_{nn}$ in fm. The calculations 
             are based on the CD-Bonn 2000 potential 
             modified by the strength factor in the $^1$S$_0$ partial 
             wave and the TM-TNI.}
    \label{tab:scatt-csb}
  \end{center}
\end{table}

\begin{figure}

\begin{center}
\psfig{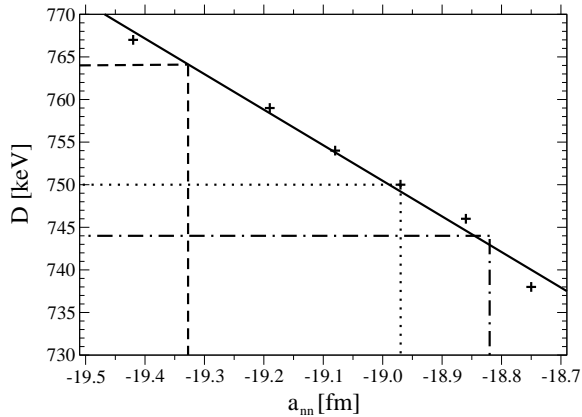}
\end{center}

\caption{Difference of the $^3$He and $^3$H BE's $D$ 
         dependent on the $nn$ scattering lengths $a_{nn}$. The crosses 
         are based on the calculations shown in Table~\ref{tab:scatt-csb} and 
         the solid line is a linear fit to the crosses.
         The dashed, the dotted and the dashed dotted lines
         mark pairs of $D$ and $a_{nn}$, which belong to the 
         experimental $D$ ($a_{nn}$ is an estimation based on the linear fit 
         in this case), the predictions of CD-Bonn~2000
         and AV18, respectively. }

\label{fig:scatt-csb}

\end{figure}

We start considering $^3$H. The calculations 
have been done for three equal fermions using the isospin formalism.
We used an averaged nucleon mass $M$ with the value
$\hbar^2/ M = 41.471$ MeV fm$^2$ (the contribution of the 
$n$-$p$ mass difference will be given separately). 
The AV18 and AV18+UIX have been
used to calculate
the BE $B$, the mean value 
of the kinetic energy $T$ as well as the $S'$, $P$- and 
$D$-state probabilities. The results are given in Table~\ref{tab:hres}
 corresponding
to two different calculations: $i)$ for total isospin limited to
$T=1/2$ and $ii)$ including also $T=3/2$. The 
occupation probability $P_{T=3/2}$ of this state is given in the last
column of Table~\ref{tab:hres}. 
The first four rows of the table show the
Pisa group results, whereas the last four show the Bochum group results. 
In the later case, the BE  $|E|$ is determined 
from the eigenvalue spectrum of the Faddeev equations. 
Additionally, we present the absolute value of the 
expectation value of the Hamiltonian $B$. 
One sees that both values agree within the numerical accuracy.
In Table~\ref{tab:heres} the same set of results are given for $^3$He.
For the $T=3/2$ calculations we find good agreement for the BE results and the 
wave function properties for both nuclei. The BE's 
are in agreement within 2~keV or 0.05~\%. The deviations for 
the wave function properties, especially for $P_{S'}$, 
are slightly bigger, but 
remain below 0.4~\%. This is below our numerical error bounds
and confirms the reliability of both methods, even in presence 
of a TNI. The tables also reveal a small, but appreciable, contribution of 
the $T=3/2$ state to the BE. 
Its inclusion produces 5 keV (8 keV) more binding in $^3$H ($^3$He). 
It should be noted that the $T=1/2$ results depend on the numerical 
method. The truncation of the Hilbert space to $T=1/2$ leads 
to average $pp$ ($nn$) and $np$ 
matrix elements in the isospin $t=1$ $NN$ channels.
This averaging is performed for the potential matrix elements 
in case of the Pisa calculations, but for the $t$-matrix in case 
of the Bochum scheme. The small, but visible differences show that 
benchmarks to this accuracy require the comparison of fully 
charge-dependent calculations. This inconsistency also 
shows up in a visible deviation of $|E|$ and $B$ for the 
$T=1/2$ Faddeev calculations, because $B$ is based on matrix 
elements of the potential wheareas $|E|$ is based on the 
$t$-matrix.   

The contribution of the n-p mass difference is visible, but sufficiently 
small to be treated perturbatively. 
Therefore we show only perturbative estimates in Table~\ref{tab:pertmass}. 
The positive
sign in the tritium case indicates a slightly more bound system,
conversely the $^3$He results slightly less bound. Again we find 
an encouraging agreement between the Pisa and Bochum results.

Taking into account 
the contribution of the n-p mass difference and averaging the
Pisa and Bochum results, the final values
of the BE's for the AV18+UIX are: 
$B(^3{\rm H})=8.485(2)$ MeV and $B(^3{\rm He})=7.741(2)$ MeV. 
This is to be compared to the experimental
values: $B_{exp}(^3{\rm H})=8.482$~MeV and $B_{exp}(^3{\rm He})=7.718$~MeV. 
Therefore, the AV18+UIX potential overbinds the tritium only by 3~keV,
whereas  the $^3$He is overbound by 23~keV.
This can be better analyzed looking at the predicted 
BE difference $D=744$~keV, 
which is 20~keV smaller than the experimental value.

The calculations have been performed using the static Coulomb
potential, i.e. the momentum dependence through $\alpha'$ has
been disregarded~\cite{austen83}, so that our results correspond to 
$\alpha'=\alpha$. It has been estimated that 
the momentum dependence of $\alpha$ might contribute 
about half of the $20$~keV shortage found in the value of 
$D$~\cite{pieperpriv,wu93}. 
Therefore our results show that there is room from other 
contributions as for example CSB of TNI terms and a refinement of
relativistic effects on $D$.

The contributions to $D$ of different parts of the interaction 
have been studied calculating
the $^3$H and $^3$He BE's omitting these parts
and comparing to the full calculations.
Note that this is not perturbative.
The results for the 
AV18+UIX potential including isospin states $T=1/2$ and $3/2$
states are collected in Table~\ref{tab:csbcontr}. 
We distinguish 
$i)$ the nuclear CSB terms,
$ii)$ the point Coulomb interaction, 
$iii)$ the complete $pp$ and $np$ Coulomb interaction, which includes
the finite size charge distributions, the one- and two-photon terms and the 
Darwin-Foldy and vacuum polarization interactions,
$iv)$ the magnetic moment interaction, $v)$ the $n$-$p$ mass difference.

One sees that the electromagnetic interaction on top 
of point Coulomb visibly contributes 
to $D$ and cannot be neglected. This raises the interesting question,
whether CD-Bonn~2000, coming without an elaborate electromagnetic force,
can also describe $D$. 

To this aim we performed $3N$ BE calculations using the 
CD-Bonn~2000 interaction. The results are given in Table~\ref{tab:3nbound}. 
Again, the $NN$ interaction underbinds the $3N$ nuclei. Therefore we 
augmented the Hamiltonian by the Tucson-Melbourne (TM) TNI 
\cite{coon79,coon81,coon01}. The strength of the original model 
has been adjusted to reproduce the experimental $^3$H BE
as described in \cite{nogga01d}. It results the $\pi NN$ cut-off value 
$\Lambda = 4.795$~m$_\pi$. Again $^3$He is overbound. The mass
difference $D$, of $750$~keV
is slightly improved compared to our result for AV18 and UIX.

In Ref.~\cite{machleidt01b} it has been observed that $D$ is only sensitive 
to CSB in the $S$-wave. Unfortunately, $nn$ scattering 
is only poorly known experimentally. There is only one datum for 
the scattering length, which is still controversal
\cite{gabioud79,gonzalez99,schori87,slaus89,howell98,huhn00a,huhn00b,miller90}. 
AV18 is adjusted to $a_{nn}=-18.82$~fm, whereas for CD-Bonn~2000 $a_{nn}$ 
results in $-18.97$~fm. Its charge-dependence is based on 
theoretical predictions of the full Bonn model \cite{machleidt01a}. 
To pin down the origin of the difference of 
the predictions of both models,  we modified the $^1$S$_0$ $nn$ 
interaction of CD-Bonn~2000 by a factor $\lambda$ and calculated the resulting 
$nn$ scattering length $a_{nn}$, the $^3$H BE and $D$. 
We found a strong linear correlation 
of $a_{nn}$ and $D$ shown in Table~\ref{tab:scatt-csb} 
and Fig.~\ref{fig:scatt-csb}. Moreover, the prediction of AV18+UIX 
perfectly fit into the results obtained from CD-Bonn~2000 and TM
(see the dashed-dotted marks in the figure). This shows that the 
dependence of $D$ on the interaction can be traced back to 
different predictions for the $nn$ scattering length. The very different 
treatment of electromagnetic interactions and the differences of the CSB 
in higher partial waves do not affect appreciable $D$. The deviation of the 
$NN$+TNI force prediction of $D$ to the experimental value might be 
caused, as stated before, by CSB TNI terms not considered in the present
description or by relativistic effects.
As a consequence
we cannot estimate $a_{nn}$ from our calculations of $D$. However, we would 
like to note that the $a_{nn} \approx -16.3$~fm 
found in Refs.~\cite{huhn00a,huhn00b} would worsen our description of the 
$3N$ BE difference significantly.

In summary we have calculated the $^3$H and $^3$He BE's based on modern 
$NN$ interaction models including TNI terms using two different 
numerical methods. Our results showed the stability 
and reliablity of both schemes. Using only $NN$ forces the BE's 
are too small, calling for TNI terms. These lead by construction 
to the experimental $^3$H BE. We found that the BE difference 
of $^3$H and $^3$He is predicted nearly model independently. 
We could trace back the remaining model sensitivity to the 
differences in the predictions for the $nn$ scattering length. 
However, uncertainties arising from CSB TNI terms and relativity 
do not allow us to extract the scattering length from 
the $^3$H and $^3$He BE difference. The model dependence 
arising from the different electromagnetic parts of the interactions
is extremely small.

A.N. acknowledge partial support from 
NSF grant\# PHY0070858. Parts of the numerical calculations 
have been performed on the Cray T3E of the NIC in J\"ulich,
Germany. 

\bibliography{literatur}

\begin{thebibliography}{31}
\expandafter\ifx\csname natexlab\endcsname\relax\def\natexlab#1{#1}\fi
\expandafter\ifx\csname bibnamefont\endcsname\relax
  \def\bibnamefont#1{#1}\fi
\expandafter\ifx\csname bibfnamefont\endcsname\relax
  \def\bibfnamefont#1{#1}\fi
\expandafter\ifx\csname citenamefont\endcsname\relax
  \def\citenamefont#1{#1}\fi
\expandafter\ifx\csname url\endcsname\relax
  \def\url#1{\texttt{#1}}\fi
\expandafter\ifx\csname urlprefix\endcsname\relax\def\urlprefix{URL }\fi
\providecommand{\bibinfo}[2]{#2}
\providecommand{\eprint}[2][]{\url{#2}}

\bibitem[{\citenamefont{{R. Machleidt} et~al.}(1996)\citenamefont{{R.
  Machleidt}, {F.~Sammarruca}, and {Y.~Song}}}]{machleidt96}
\bibinfo{author}{\bibnamefont{{R. Machleidt}}},
  \bibinfo{author}{\bibnamefont{{F.~Sammarruca}}}, \bibnamefont{and}
  \bibinfo{author}{\bibnamefont{{Y.~Song}}}, \bibinfo{journal}{Phys. Rev. C}
  \textbf{\bibinfo{volume}{53}}, \bibinfo{pages}{R1483} (\bibinfo{year}{1996}).

\bibitem[{\citenamefont{{R.B. Wiringa} et~al.}(1995)\citenamefont{{R.B.
  Wiringa}, {V.G.J. Stoks}, and {R. Schiavilla}}}]{wiringa95}
\bibinfo{author}{\bibnamefont{{R.B. Wiringa}}},
  \bibinfo{author}{\bibnamefont{{V.G.J. Stoks}}}, \bibnamefont{and}
  \bibinfo{author}{\bibnamefont{{R. Schiavilla}}}, \bibinfo{journal}{Phys. Rev.
  C} \textbf{\bibinfo{volume}{51}}, \bibinfo{pages}{38} (\bibinfo{year}{1995}).

\bibitem[{\citenamefont{{V.G.J. Stoks} et~al.}(1994)\citenamefont{{V.G.J.
  Stoks}, {R.A.M. Klomp}, {C.P.F. Terheggen}, and {J.J. de Swart}}}]{stoks94}
\bibinfo{author}{\bibnamefont{{V.G.J. Stoks}}},
  \bibinfo{author}{\bibnamefont{{R.A.M. Klomp}}},
  \bibinfo{author}{\bibnamefont{{C.P.F. Terheggen}}}, \bibnamefont{and}
  \bibinfo{author}{\bibnamefont{{J.J. de Swart}}}, \bibinfo{journal}{Phys. Rev.
  C} \textbf{\bibinfo{volume}{49}}, \bibinfo{pages}{2950}
  (\bibinfo{year}{1994}).

\bibitem[{\citenamefont{{R. Machleidt}}(2001)}]{machleidt01a}
\bibinfo{author}{\bibnamefont{{R. Machleidt}}}, \bibinfo{journal}{Phys. Rev. C}
  \textbf{\bibinfo{volume}{63}}, \bibinfo{pages}{024001}
  (\bibinfo{year}{2001}).

\bibitem[{\citenamefont{{ {A. Nogga}, {H. Kamada}, {W.
  Gl\"ockle}}}(2000)}]{nogga00}
\bibinfo{author}{\bibnamefont{{ {A. Nogga}, {H. Kamada}, {W. Gl\"ockle}}}},
  \bibinfo{journal}{Phys. Rev. Lett.} \textbf{\bibinfo{volume}{85}},
  \bibinfo{pages}{944} (\bibinfo{year}{2000}).

\bibitem[{\citenamefont{{J. Carlson, and R.Schiavilla}}(1998)}]{carlson98}
\bibinfo{author}{\bibnamefont{{J. Carlson and R.Schiavilla}}},
  \bibinfo{journal}{Rev. Mod. Phys.} \textbf{\bibinfo{volume}{70}},
  \bibinfo{pages}{743} (\bibinfo{year}{1998}).

\bibitem[{\citenamefont{{B.S. Pudliner} et~al.}(1995)\citenamefont{{B.S.
  Pudliner}, {V.R. Pandharipande}, {J. Carlson}, and {R.B.
  Wiringa}}}]{pudliner95}
\bibinfo{author}{\bibnamefont{{B.S. Pudliner}}},
  \bibinfo{author}{\bibnamefont{{V.R. Pandharipande}}},
  \bibinfo{author}{\bibnamefont{{J. Carlson}}}, \bibnamefont{and}
  \bibinfo{author}{\bibnamefont{{R.B. Wiringa}}}, \bibinfo{journal}{Phys. Rev.
  Lett.} \textbf{\bibinfo{volume}{74}}, \bibinfo{pages}{4396}
  (\bibinfo{year}{1995}).

\bibitem[{\citenamefont{{Steven C. Pieper,V.R. Pandharipande, R.B. Wiringa, J.
  Carlson }}(2001)}]{pieper01}
\bibinfo{author}{\bibnamefont{{S.C. Pieper,V.R. Pandharipande, R.B.
  Wiringa, and J. Carlson }}}, \bibinfo{journal}{Phys. Rev. C}
  \textbf{\bibinfo{volume}{64}}, \bibinfo{pages}{014001}
  (\bibinfo{year}{2001}).

\bibitem[{\citenamefont{{J.L. Friar} et~al.}(1987)\citenamefont{{J.L. Friar}, {
  B. F. Gibson}, and { G.L. Payne}}}]{friar87}
\bibinfo{author}{\bibnamefont{{J.L. Friar}}}, \bibinfo{author}{\bibnamefont{{
  B. F. Gibson}}}, \bibnamefont{and} \bibinfo{author}{\bibnamefont{{ G.L.
  Payne}}}, \bibinfo{journal}{Phys. Rev. C} \textbf{\bibinfo{volume}{35}},
  \bibinfo{pages}{1502} (\bibinfo{year}{1987}).

\bibitem[{\citenamefont{{Y. Wu} et~al.}(1990)\citenamefont{{Y. Wu}, {S.
  Ishikawa}, and {T. Sasakawa}}}]{wu90}
\bibinfo{author}{\bibnamefont{{Y. Wu}}}, \bibinfo{author}{\bibnamefont{{S.
  Ishikawa}}}, \bibnamefont{and} \bibinfo{author}{\bibnamefont{{T. Sasakawa}}},
  \bibinfo{journal}{Phys. Rev. Lett.} \textbf{\bibinfo{volume}{64}},
  \bibinfo{pages}{1875} (\bibinfo{year}{1990}).

\bibitem[{\citenamefont{{Y. Wu} et~al.}(1993)\citenamefont{{Y. Wu}, {S.
  Ishikawa}, and {T. Sasakawa}}}]{wu93}
\bibinfo{author}{\bibnamefont{{Y. Wu}}}, \bibinfo{author}{\bibnamefont{{S.
  Ishikawa}}}, \bibnamefont{and} \bibinfo{author}{\bibnamefont{{T. Sasakawa}}},
  \bibinfo{journal}{Few-Body Systems} \textbf{\bibinfo{volume}{15}},
  \bibinfo{pages}{145} (\bibinfo{year}{1993}).

\bibitem[{\citenamefont{{A. Kievsky} et~al.}(1993)\citenamefont{{A. Kievsky},
  {M. Viviani}, and {S. Rosati}}}]{kievsky93}
\bibinfo{author}{\bibnamefont{{A. Kievsky}}}, \bibinfo{author}{\bibnamefont{{M.
  Viviani}}}, \bibnamefont{and} \bibinfo{author}{\bibnamefont{{S. Rosati}}},
  \bibinfo{journal}{Nucl. Phys.} \textbf{\bibinfo{volume}{A551}},
  \bibinfo{pages}{241} (\bibinfo{year}{1993}).

\bibitem[{\citenamefont{{A. Kievsky}}(1997)}]{kievsky97}
\bibinfo{author}{\bibnamefont{{A. Kievsky}}}, \bibinfo{journal}{Nucl. Phys.}
  \textbf{\bibinfo{volume}{A624}}, \bibinfo{pages}{125} (\bibinfo{year}{1997}).

\bibitem[{\citenamefont{{Y. Suzuki, K. Varga}}(1998)}]{suzukilec}
\bibinfo{author}{\bibnamefont{{Y. Suzuki and K. Varga}}},
  \emph{\bibinfo{title}{Stochastical variational approach to Quantum-Mechanical
  Few-Body Problems}}, vol. \bibinfo{volume}{m54} of
  \emph{\bibinfo{series}{Lecture Notes in Physics}}
  (\bibinfo{publisher}{Springer-Verlag}, \bibinfo{address}{Berlin},
  \bibinfo{year}{1998}).

\bibitem[{\citenamefont{{H. Kameyama, M. Kamimura, and Y.
  Fukushima}}(1989)}]{kameyama89}
\bibinfo{author}{\bibnamefont{{H. Kameyama, M. Kamimura, and Y. Fukushima}}},
  \bibinfo{journal}{Phys. Rev. C} \textbf{\bibinfo{volume}{40}},
  \bibinfo{pages}{974} (\bibinfo{year}{1989}).

\bibitem[{\citenamefont{{P. Navr\'atil, B.R. Barrett}}(1998)}]{navratil98}
\bibinfo{author}{\bibnamefont{{P. Navr\'atil and B.R. Barrett}}},
  \bibinfo{journal}{Phys. Rev. C} \textbf{\bibinfo{volume}{57}},
  \bibinfo{pages}{562} (\bibinfo{year}{1998}).

\bibitem[{\citenamefont{{G.J.M. Austen} and {J.J. de Swart}}(1983)}]{austen83}
\bibinfo{author}{\bibnamefont{{G.J.M. Austen}}} \bibnamefont{and}
  \bibinfo{author}{\bibnamefont{{J.J. de Swart}}}, \bibinfo{journal}{Phys. Rev.
  Lett.} \textbf{\bibinfo{volume}{50}}, \bibinfo{pages}{2039}
  (\bibinfo{year}{1983}).

\bibitem[{\citenamefont{{S.C. Pieper}}()}]{pieperpriv}
\bibinfo{author}{\bibnamefont{{S.C. Pieper}}}, \bibinfo{note}{private
  communication}.

\bibitem[{\citenamefont{{S.A. Coon, M.D. Scadron, P.C. McNamee, B.R. Barrett,
  D.W.E. Blatt, and B.H.J. McKellar}}(1979)}]{coon79}
\bibinfo{author}{\bibnamefont{{S.A. Coon, M.D. Scadron, P.C. McNamee, B.R.
  Barrett, D.W.E. Blatt, and B.H.J. McKellar}}}, \bibinfo{journal}{Nucl. Phys.}
  \textbf{\bibinfo{volume}{\bf A317}}, \bibinfo{pages}{242}
  (\bibinfo{year}{1979}).

\bibitem[{\citenamefont{{S.A. Coon} and {W. Gl\"ockle}}(1981)}]{coon81}
\bibinfo{author}{\bibnamefont{{S.A. Coon}}} \bibnamefont{and}
  \bibinfo{author}{\bibnamefont{{W. Gl\"ockle}}}, \bibinfo{journal}{Phys. Rev.
  C} \textbf{\bibinfo{volume}{23}}, \bibinfo{pages}{1790}
  (\bibinfo{year}{1981}).

\bibitem[{\citenamefont{{S.A. Coon} and {H.K. Han}}(2001)}]{coon01}
\bibinfo{author}{\bibnamefont{{S.A. Coon}}} \bibnamefont{and}
  \bibinfo{author}{\bibnamefont{{H.K. Han}}}, \bibinfo{journal}{Few-Body
  Systems} \textbf{\bibinfo{volume}{30}}, \bibinfo{pages}{131}
  (\bibinfo{year}{2001}).

\bibitem[{\citenamefont{{ {A. Nogga}, {H. Kamada}, {W. Gl\"ockle}, {B.R.
  Barrett}}}()}]{nogga01d}
\bibinfo{author}{\bibnamefont{{ {A. Nogga}, {H. Kamada}, {W. Gl\"ockle}, 
  and {B.R. Barrett}}}}, \eprint{nucl-th/0112026}.

\bibitem[{\citenamefont{{R. Machleidt} and {H. M\"uther}}(2001)}]{machleidt01b}
\bibinfo{author}{\bibnamefont{{R. Machleidt}}} \bibnamefont{and}
  \bibinfo{author}{\bibnamefont{{H. M\"uther}}}, \bibinfo{journal}{Phys. Rev.
  C} \textbf{\bibinfo{volume}{63}}, \bibinfo{pages}{034005}
  (\bibinfo{year}{2001}).

\bibitem[{\citenamefont{{B. Gabioud {\it et. al.} }}(1979)}]{gabioud79}
\bibinfo{author}{\bibnamefont{{B. Gabioud {\it et. al.} }}},
  \bibinfo{journal}{Phys. Rev. Lett.} \textbf{\bibinfo{volume}{42}},
  \bibinfo{pages}{1508} (\bibinfo{year}{1979}).

\bibitem[{\citenamefont{{D.E. Gonz\'{a}lez Trotter {\it et. al.}
  }}(1999)}]{gonzalez99}
\bibinfo{author}{\bibnamefont{{D.E. Gonz\'{a}lez Trotter {\it et. al.} }}},
  \bibinfo{journal}{Phys. Rev. Lett.} \textbf{\bibinfo{volume}{83}},
  \bibinfo{pages}{3788} (\bibinfo{year}{1999}).

\bibitem[{\citenamefont{{O. Schori {\it et. al.}}}(1987)}]{schori87}
\bibinfo{author}{\bibnamefont{{O. Schori {\it et. al.}}}},
  \bibinfo{journal}{Phys. Rev. C} \textbf{\bibinfo{volume}{35}},
  \bibinfo{pages}{2252} (\bibinfo{year}{1987}).

\bibitem[{\citenamefont{{I. \v{S}laus, Y. Akaishi, H. Tanaka
  }}(1989)}]{slaus89}
\bibinfo{author}{\bibnamefont{{I. \v{S}laus, Y. Akaishi, and H. Tanaka }}},
  \bibinfo{journal}{Phys. Rep.} \textbf{\bibinfo{volume}{173}},
  \bibinfo{pages}{259} (\bibinfo{year}{1989}).

\bibitem[{\citenamefont{{C. R. Howell {\it et al.}}}(1998)}]{howell98}
\bibinfo{author}{\bibnamefont{{C. R. Howell {\it et al.}}}},
  \bibinfo{journal}{Phys. Lett.} \textbf{\bibinfo{volume}{B444}},
  \bibinfo{pages}{252} (\bibinfo{year}{1998}).

\bibitem[{\citenamefont{{{V. Huhn}, {L. W\"atzold}, {Ch. Weber} ,{A. Siepe},{W.
  von Witsch}, {H. Wita{\l}a}, {W. Gl\"ockle} }}(2000{\natexlab{a}})}]{huhn00a}
\bibinfo{author}{\bibnamefont{{{V. Huhn}, {L. W\"atzold}, {Ch. Weber}, {A.
  Siepe}, {W. von Witsch}, {H. Wita{\l}a}, and {W. Gl\"ockle} }}},
  \bibinfo{journal}{Phys. Rev. C} \textbf{\bibinfo{volume}{63}},
  \bibinfo{pages}{014003} (\bibinfo{year}{2000}{\natexlab{a}}).

\bibitem[{\citenamefont{{{V. Huhn}, {L. W\"atzold}, {Ch. Weber} ,{A. Siepe},{W.
  von Witsch}, {H. Wita{\l}a}, {W. Gl\"ockle} }}(2000{\natexlab{b}})}]{huhn00b}
\bibinfo{author}{\bibnamefont{{{V. Huhn}, {L. W\"atzold}, {Ch. Weber} , {A.
  Siepe}, {W. von Witsch}, {H. Wita{\l}a}, and {W. Gl\"ockle} }}},
  \bibinfo{journal}{Phys. Rev. Lett.} \textbf{\bibinfo{volume}{85}},
  \bibinfo{pages}{1190} (\bibinfo{year}{2000}{\natexlab{b}}).

\bibitem[{\citenamefont{{G.A. Miller, B.M.K. Nefkens, and I.
  \v{S}laus}}(1990)}]{miller90}
\bibinfo{author}{\bibnamefont{{G.A. Miller, B.M.K. Nefkens, and I.
  \v{S}laus}}}, \bibinfo{journal}{Phys. Rep.} \textbf{\bibinfo{volume}{194}},
  \bibinfo{pages}{1} (\bibinfo{year}{1990}).

\end{thebibliography}

\end{document}